\documentclass[fleqn,usenatbib]{mnras}

\usepackage{newtxtext,newtxmath}

\usepackage[T1]{fontenc}

\DeclareRobustCommand{\VAN}[3]{#2}
\let\VANthebibliography\thebibliography
\def\thebibliography{\DeclareRobustCommand{\VAN}[3]{##3}\VANthebibliography}

\usepackage{graphicx}	
\usepackage{amsmath}	

\title[Jet Width Profile of CTA 102]{Investigating the Jet Width Profile of CTA 102 with Very Long Baseline Interferometry at Parsec Scales}

\author[Z. H. Ng et al.]{
Z. H. Ng,$^{1}$
J. C. Algaba$^{1}$ \thanks{corresponding author, e-mail: algaba@um.edu.my}
Z. Z. Abidin,$^{1}$
\\
$^{1}$Centre for Astronomy \& Astrophysics Research, Department of Physics, Faculty of Science, Universiti Malaya, 50603, Kuala Lumpur, Malaysia\\
}

\date{Accepted XXX. Received YYY; in original form ZZZ}

\pubyear{2025}

\begin{document}
\label{firstpage}
\pagerange{\pageref{firstpage}--\pageref{lastpage}}
\maketitle

\begin{abstract}
Active Galactic Nucleus jets have long be thought to exhibit a conical jet shape, but recently, several jets were found to have a transition from parabolic to conical structure. As more sources are investigated, this collimation profile appears to represent a common paradigm. Previous works suggest that the Bondi radius may serve as an indicator of the transition location, although discrepancies have been observed in some sources. To explore this further, we selected CTA~102 for which existing literature presents mixed evidence regarding the presence of a jet geometry break. We investigated the jet width profile of CTA~102 to study the possible transition changes in the jet, thereby improving the understanding of connection between Bondi radius and jet transition. We used multi-frequency VLBA images of CTA~102 at 2, 5, and 8 (single epoch), and 15, 22 and 43~GHz (stacked). The jet width profile was modeled with a single power law $W_{\rm{jet}} \propto r^{\epsilon}$ yielding a power-law index of $\epsilon = 0.69\pm0.02$, indicative of a quasi-parabolic geometry with no clear transition to a conical regime. The absence of discernible structural break around the Bondi radius implies that the physical conditions associated with the radius alone are insufficient to explain the jet collimation behaviour. On the other hand, we observe oscillatory features in the jet width profile, suggesting the influence of additional physical processes beyond gravitational confinement. These findings contribute to a more nuanced understanding of jet collimation in AGN and highlight the complexity of jet-environment interactions.
\end{abstract}

\begin{keywords}
galaxies: jets -- galaxies: active -- techniques: interferometric
\end{keywords}


\section{Introduction}

Active Galactic Nuclei (AGNs) are a type of highly luminous and highly variable astronomical object at the centre of some galaxies. It is believed that the central engine is powered by a supermassive black hole (SMBH) around $10^7-10^{10} M_{\odot}$ \citep{greenstein1979quasi}, with accretion matter around the centre and matter accreting into it. AGNs can be generally classified into two groups: radio loud or radio quiet, based on their relative radio emission intensity compared to their optical intensity \citep{strittmatter1980radio, kellermann1989vla, xiao2022radio}. Another common feature is that Radio-Loud AGN produce powerful relativistic jets featuring synchrotron emission whereas Radio-Quiet AGNs have weak, or no jet produced, and are typically attributed to thermal radiation. There are about 10 percent of radio-loud AGNs, and within them, different characteristics had led to different categorisation \citep{urry1995unified}. One of the on-going debate suggests that these radio-loud AGNs are the same type but only with different viewing angle. 

One of the most prominent features of AGNs is the relativistic jet. These collimated relativistic plasma outflows are believed to come from the central engine of the galactic nuclei. Two mechanisms, Blandford-Znajek (BZ) \citep{blandford1977electromagnetic} and Blandford-Payne (BP) \citep{blandford1982hydromagnetic} were proposed to explain the origin of the jet. BZ process suggested that the matter from the accretion disk extract the rotational energy of a black hole, where BP process proposed that these matter extract energy from the magnetosphere. While the precise mechanism remains an active area of research, both proposed models posit that the energy is converted from the magnetic field into kinetic energy in the form of a Poynting flux, leading to the acceleration of matter from the vicinity of the AGN, collimating and forming a relativistic superluminating plasma jet.

In some models, both jet acceleration and collimation are posited to occur concurrently near the base of the relativistic jet, commonly referred to as the acceleration and collimation zone (ACZ) via the general relativistic magnetohydrodynamic (GRMHD) effects \citep{nakamura2013parabolic}. Therefore, studying the jet collimation therefore becomes a hot topic in the field. From previous study, it is found that a transition of parabolic to conical jet \citep{asada2012structure} is a common feature in many nearby sources \citep{kovalev2020transition}, for instance M87 \citep{asada2012structure, nakamura2018parabolic}, NGC~4261 \citep{nakahara2018finding, nakahara2019two, yan2023kinematics}, NGC~6251 \citep{tseng2016structural} to mention some. Distant galaxies on the other hand, were observed to have only conical jet structure \citep{kovalev2020transition}.

In the case of M87, \citet{asada2012structure} observed that the parabolic collimation of the jet occuring around the region of Bondi radius. This radius signifies the boundary the SMBH dominant gravitational influence, ranging  from $10^5$ to $10^6$ Schwarzschild radius ($R_{\rm{s}}$). Beyond this region, it is hypothesized that the jet enter a regime where its expansion is shaped by the interplay between the external interstellar medium (ISM) pressure and the jet pressure, resulting to a transition into a conical free expansion. The Bondi radius, $R_{\rm{B}}$, calculated as
\begin{equation}
	R_{\rm{B}}=\frac{2GM_{\rm{BH}}}{c_{\rm{s}}^2},
\end{equation}
where $c_{\rm{s}}= \sqrt{\frac{\gamma k_{\rm{B}}T}{\mu m_{\rm{p}}}}$, typically $\mu$ = 0.6, the mean molecular weight, $\gamma = \frac{5}{3}$  is the adiabatic index of the accreting gas, and $m_{\rm{p}}$ is the proton mass. On the other hand, the sphere of gravitational influence (SGI), generally defines the region where the gravitational pull of a central mass dominates over the gravitational influence of other. Although conceptually distinct, the SGI typically yields the same order of magnitude as the Bondi radius and thus provides an alternative framework for examining the location of jet width transitions. This equivalence arises because both radii characterise scales where the SMBH's gravity dominates over the external influences \citep{russell2015inside}. The SGI is calculated as 
\begin{equation}
	R_{\rm{SGI}} = \frac{GM_{\rm {BH}}}{\sigma^2},
\end{equation}
where $\sigma$ is the stellar velocity dispersion. 

Previous studies on M87 \citep{asada2012structure} showed that the transition occurs around $2.5 \times 10^5 R_{\rm{s}}$, close to the Bondi radius $3.8 \times 10^5 R_s$. Studies on NGC~6251 by \citet{tseng2016structural} showed that the transition located around $(1-2) \times 10^5 R_{\rm{s}}$ is also close to the SGI ($\sim5 \times 10^5 R_{\rm{s}}$). However, this transition location close to the Bondi radius or SGI is not a universal case as some AGNs were found to have their transition located far away from it and quite a number of AGNs not showing any transition at this distance. The jet collimation profile of NGC~4261 (Bondi radius around $6.9 \times 10^5 R_{\rm{s}}$) reported by \citet{nakahara2018finding} showed that the transition located at $\sim10^4 R_{\rm{s}}$. \citet{okino2022collimation} also showed that the collimation of quasar 3C~273 located nearly several-to-ten times further than the estimated SGI location. Their figure 12 showed that the transition of the jet collimation for several sources deviates from the sources respective locations derived from $M_{\rm{BH}} - \sigma_{\rm{d}} $ relation presenting that the collimation may not be governed only by the SGI of the central engine. Hence, more AGN jets need to be studied to determine if this is a general paradigm or if these sources are outliers from the trend.

CTA 102 (B2230+114) was first observed by \citet{harris1960radio} at a redshift of 1.037 \citep{schmidt1965large}. It is a type of highly polarized quasar with a linear optical polarization above 3 percent \citep{veron2003catalogue}. Its central supermassive black hole mass measured to be around $10^{8.93} M_{\odot}$ \citep{zamaninasab2013evidence, fromm2015location}, giving a Schwarzschild radius of approximately  $8.193\times 10^{-5}$ pc. It exhibits a relativistic jet with a viewing angle of $2.6^\circ$ \citep{fromm2013catchingIII, jorstad2005polarimetric} extended towards the southeast spanning a projected distance up to 25 mas (deprojected 550 mas or 4.5 kpc) \citep{fromm2013catchingII}. This conversion indicates that 1 mas is equivalent to 8.11 pc, while 1 projected mas corresponds to 22 deprojected mas. The jet showed a quasi-parabolic jet collimation profile (power index, $\epsilon = 0.80 \pm 0.03$ \citep{pushkarev2017mojave} with no jet break observed around the Bondi radius. Results from \citet{fromm2013catchingIII} also showed an $\epsilon$ value of $0.80 \pm 0.1$. However, \citet{algaba2017resolving} proposed that there may be a possible jet break at $\sim$30 to 100 pc deprojected ($\sim10^5$ to $10^6$ $R_{\rm{s}}$) by using VLBI core observations at 1.6 GHz to 86 GHz. 

Building upon the earlier work by \citep{fromm2011catchingI, fromm2013catchingII, fromm2013catchingIII}, where the jet width profile was not extensively analyzed, and motivated by the proposal from \citep{algaba2017resolving} regarding a possible jet break, this study aims to re-investigate the jet width profile of CTA 102 to gain a deeper insight into its structural features. Additionally, we investigate the viability of a helical model in describing the jet streamline, analyse the correlation between the Bondi radius with the jet transition across multiple sources and identify the jet humps feature from M87, 3C~264, BL Lac and 3C 273 to be a possible common feature among AGN sources.
 
This paper is structured as follows: In section 2, we present the observations and data reduction method. In section 3, we present the results of the jet image and jet width profile. In section 4, we present the discussion on the jet width structure, and in section 5 we summarize our work. We adopt a cosmology of $\Omega_{\rm{m}} = 0.27, \Omega_{\Lambda} = 0.73$ and $H_{0}= 71 \rm{km s}^{-1} \rm{Mpc}^{-1}$

\section{Observations and Data Reduction}
\subsection{Observations}
For the case of CTA 102, the Bondi radius, once converted into observable distances, is expected to be located at around 11.3 mas ($\sim10^6 R_{\rm{s}}$) from the core, therefore the VLBI technique, which can provide an angular resolution up to sub-milliarcsecond which is essential at analysing the jet transition around the Bondi Radius. To have full coverage of the VLBI jet of CTA102 on these scales, we collected VLBA datasets at 2, 5, 8, 15 and 43 GHz. We used stacked images whenever improving the signal-to-noise ratio (SNR) was possible, but this was not feasible for low frequencies as we did not find enough observation time for low frequency from the archive. Therefore, the low-frequency data (2, 5, and 8 GHz) were from a single epoch.

The single epoch observations were taken from National Radio Astronomy Observatory (NRAO) archive data. Realising that there is a radio flare around the year 2006 \citep{fromm2011catchingI} that likely to introduce significant variability and structural changes in the jet, such as bright knots, increased opacity or altered brightness temperature that can distort the jet width measurement \citep{fromm2013catchingIII}, we omitted data around this year. We selected project code BA064 using 2.2095 GHz and 8.1185 GHz observed on date 2003 Feb 21 with four IFs, and a total bandwidth of 32 MHz; project code V019 using 4.8159 GHz on date 1998 May 25 with two IFs, and a total bandwidth of 32 MHz observation alternated with other sources to obtain a good UV coverage. For multi-epoch data (15, 22, and 43 GHz), we took the project code BW086 using 22 GHz observing for 3 years from 2006 to 2008 with four IFs, and a total bandwidth of 32 MHz; MOJAVE (Monitoring Of Jets in Active galactic nuclei with VLBA Experiments) data \footnote {https://www.cv.nrao.edu/MOJAVE/sourcepages/2230+114.shtml}, and BU (Boston University VLBA-BU-BLAZAR Monitoring programme) data \footnote {https://www.bu.edu/blazars/VLBA\_GLAST/cta102.html} for 15 and 43 GHz respectively observing in a large time range. We did not discard the 2006 flare data for multi-epoch data as the weightage will be insignificant after we averaged out the data. The details are presented in table~\ref{tab:Observation data}. 

We applied the standard data calibration process using Astronomical Image Processing System (AIPS) software package (NRAO) \citep{wells1985nrao} for 2, 5, 8 and 22 GHz, then applied CLEAN algorithm to obtain a clean map using Difmap software package \citep{shepherd1997difmap}. The details will be discussed in the next subsection. For MOJAVE and BU data, we obtained the beam size for each epoch and then applied the restore function in Difmap, standardising the epochs into the same circular beam by taking the average of all epochs. 

\begin{figure}
	\includegraphics[width=\columnwidth]{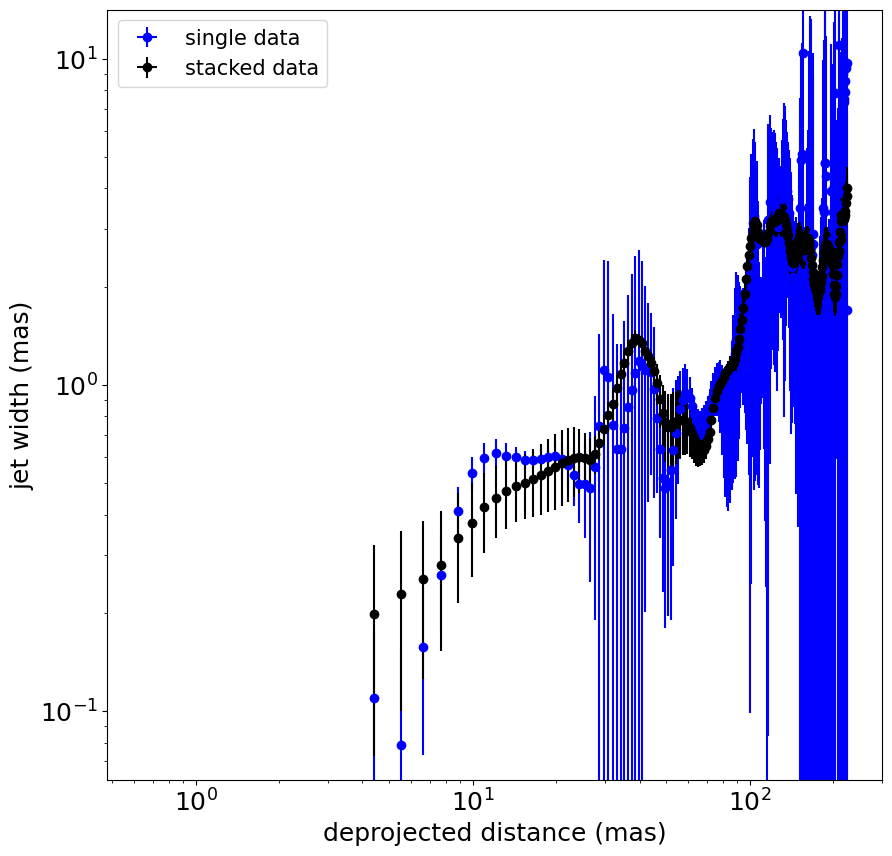}
	\caption{15 GHz Single Epoch Jet vs Multi-epoch Jet Width Profile}
    \label{fig: singlevsstack}
\end{figure}

In order to check that our analysis is not biased by the use of single epoch maps at certain frequencies and stacked maps at others, we selected a single epoch 15 GHz observation that is temporally close to the 2003 single-epoch data and generated the corresponding jet width profile. This was then compared with the jet width profile derived from the stacked 15 GHz. As shown in Fig.~\ref{fig: singlevsstack}, the differences between the two profiles are found to be insignificant, particularly in the outer regions of the jet, while simultaneously achieving an improved SNR with the stacked data. While some discrepancies are observed in the inner jet region, a small portion of the profile data near the core was excluded in the analysis due to unresolved nature. As such, they remain within the measurement uncertainties. Therefore, we consider the combined use of single epoch and stacked images to be justified and reliable for our analysis.

\begin{table}
\caption{Observation data}
\begin{center}
\begin{tabular}{|| c c c c ||} 
	\hline
	Project Code & Observation Date & N & ${\nu}$ (GHz)  \\ [0.1ex]
	\hline \hline
	BA064 & 2003-02-21 & 1 & 2.3 \\ 
	\hline
	V019 & 1998-05-25  & 1 & 4.8 \\
	\hline
	BA064 & 2003-02-21 & 1 & 8.3 \\
	\hline
	MOJAVE* & 1994-08-31 - 2015-08-20 & 42 & 15.3 \\
	\hline
	BW086 & 2006-08-03 - 2008-01-03 & 6 & 22.2 \\ 
	\hline
	 BU* & 2007-06-14 - 2015-12-05 & 98 & 43.2 \\ 
	\hline
\end{tabular}
\end{center}
\textbf{Note:} The columns are arranged as follow: project code; observation date;number of epochs; frequency. *MOJAVE (Monitoring Of Jets in Active galactic nuclei with VLBA Experiments); BU(Boston University VLBA-BU-BLAZAR Monitoring programme) 
\label{tab:Observation data}
\end{table}

\begin{table}
\caption{Image Parameters}
\begin{center}
\begin{tabular}{|| c c c c c c ||} 
	\hline
	${\nu}$ & $I_{peak}$ & ${\sigma}$ & ${\theta}_{circular}$ & IFs \\ [0.1ex]
	(GHz) & (Jy $beam^{-1}$) & (mJy $beam^{-1}$) & (mas) &  \\
	\hline \hline
	2.3 & 1.766 & 3.503 & 5.875 & 4\\ 
	\hline
	4.8 & 1.425 & 3.231 & 2.608 & 4\\
	\hline
	8.3 & 1.805 & 1.129 & 1.529 & 4\\
	\hline
	15.3 & 2.205 & 0.100 & 0.810 & 4\\
	\hline
	22.2 & 1.996 & 0.570 & 0.598 & 4\\ 
	\hline
	43.2 & 1.776 & 0.059 & 0.308 & 4\\ 
	\hline
\end{tabular}
\end{center}
\textbf{Note:} The columns are arranged as follow: frequency; peak intensity; image rms noise; circular synthesized beamsize; number of IFs
\label{tab:Image parameters}
\end{table}

\subsection{Data Reduction}
A standard calibration process using AIPS was started with loading fits file (FILTD) for the 2, 5 8 and 22 GHz data, sorting data (MSORT), creating index table (INDXR), examining data quality (LISTR, UVPLT, SNPLT), auto-correlation (ACCOR), amplitude calibration (APCAL), fringe-fitting (FRING), bandpass calibration (BPASS), plotting total and cross-power spectra (visibility amplitude and phase against frequency) (POSSM) and applying derived calibration solutions to the visibility set (CLCAL). The process ends with splitting and outsource the calibrated visibility set (SPLIT, FITTP). 

\begin{figure*}
	\includegraphics[scale=0.4]{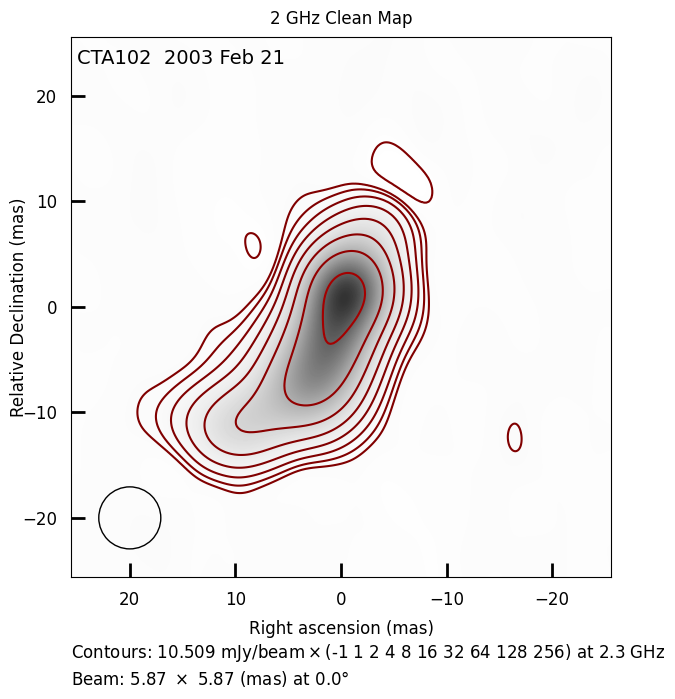}
	\includegraphics[scale=0.4]{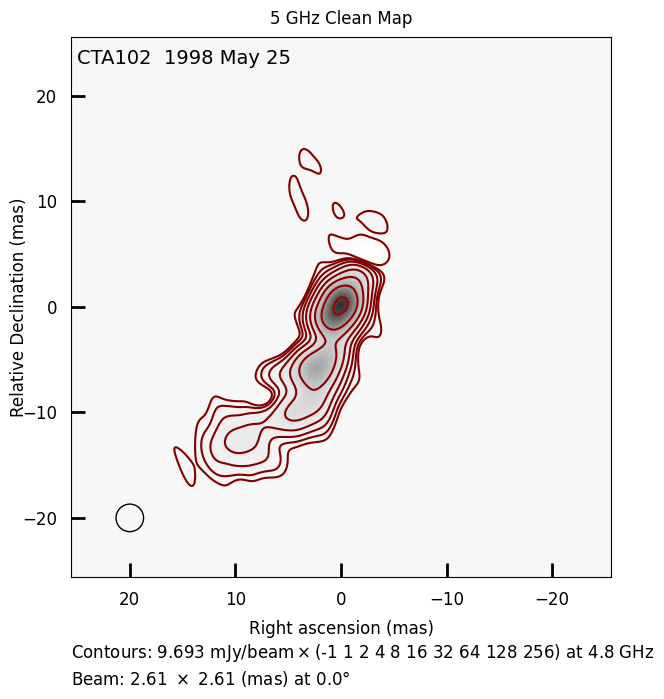}
	\includegraphics[scale=0.4]{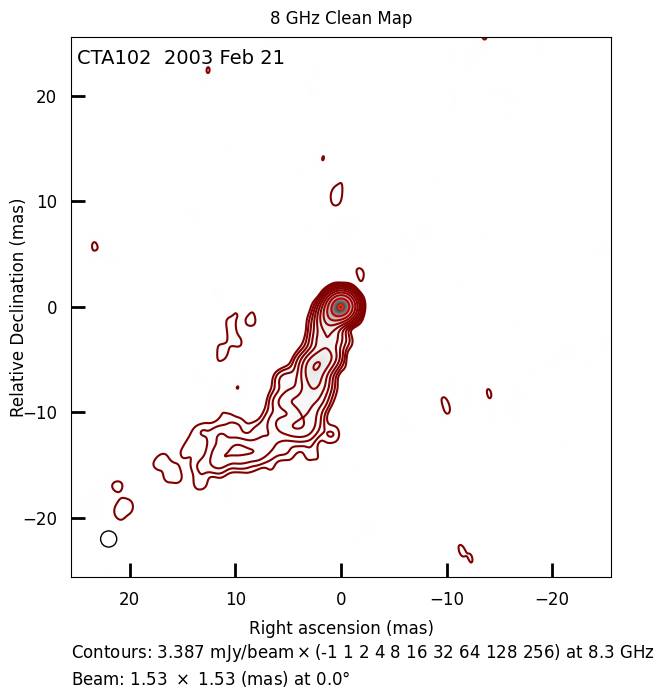}
	\includegraphics[scale=0.4]{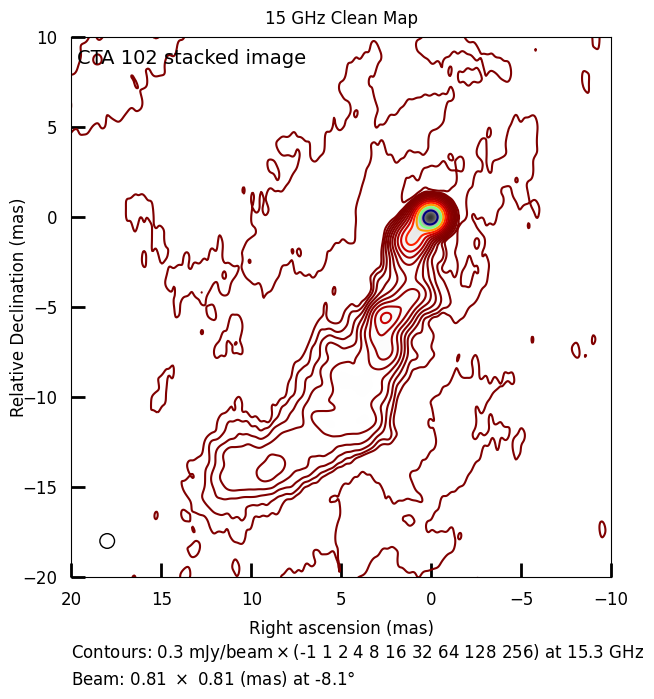}
	\includegraphics[scale=0.4]{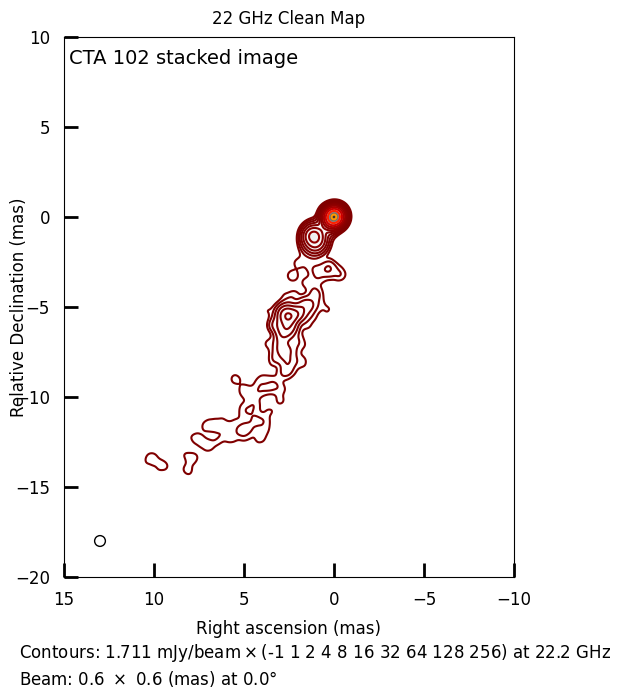}
	\includegraphics[scale=0.4]{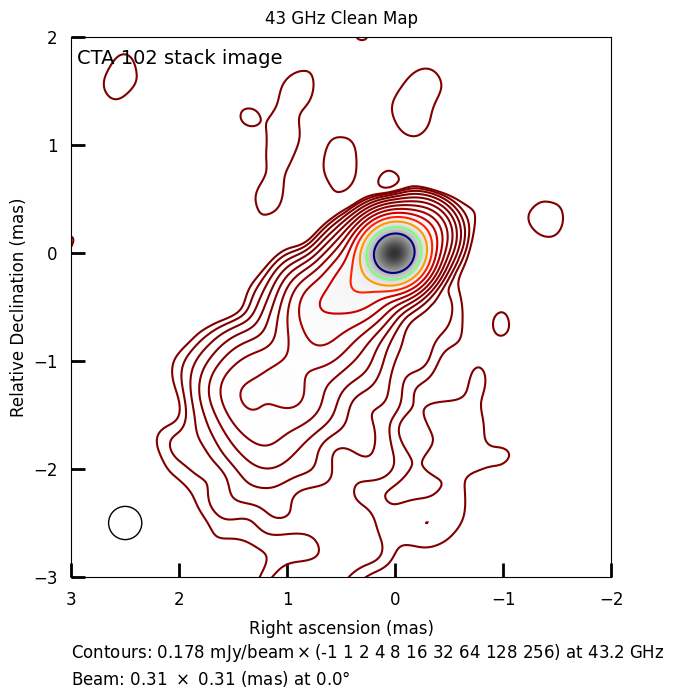}
    \caption{Multifrequencies CLEAN map}
   	 \label{fig: All Clean Map}
\end{figure*}

The data were then transferred to Difmap software package. A standard CLEAN algorithm is applied to obtain a clean map. The details of the image parameters are shown in table~\ref{tab:Image parameters}. A final clean map is then obtained with mapplot clean command at 3 times rms value. For 15 GHz multi-epoch data, we used the stacked image from MOJAVE, restored with a circular beam. For 22 and 43 GHz, we applied the weighted sum to stack the multi-epoch data and obtained a weighted rms. All CLEAN map were shown in Fig.~\ref{fig: All Clean Map}.

\begin{figure}
	\includegraphics[width=\columnwidth]{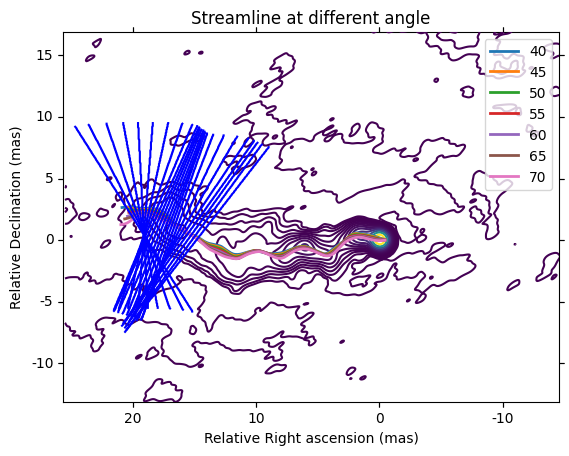}
    \caption{Representation of different streamline for differerent rotation at 15 GHz. The colours indicating different streamline obtained by rotating the 15 GHz jet streamline clockwise from $40{^\circ}-70{^\circ}$ with $5{^\circ}$ increment. The dark blue lines indicating a portion of  perpendicular slice made using the 60 degree streamline every 2 pixels from the core}
    \label{fig:streamline rotation}
\end{figure}

\begin{figure}
	\includegraphics[width=\columnwidth]{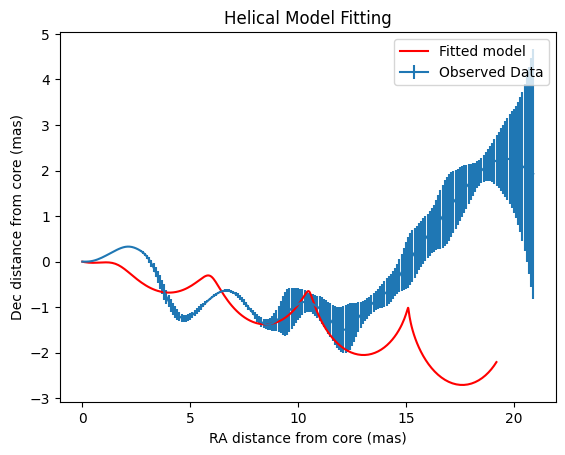}
    \caption{Helical Model Fitting for 15 GHz Jet Streamline}
    \label{fig:helical_fit}
\end{figure}

\begin{figure*}
	\includegraphics[scale=1]{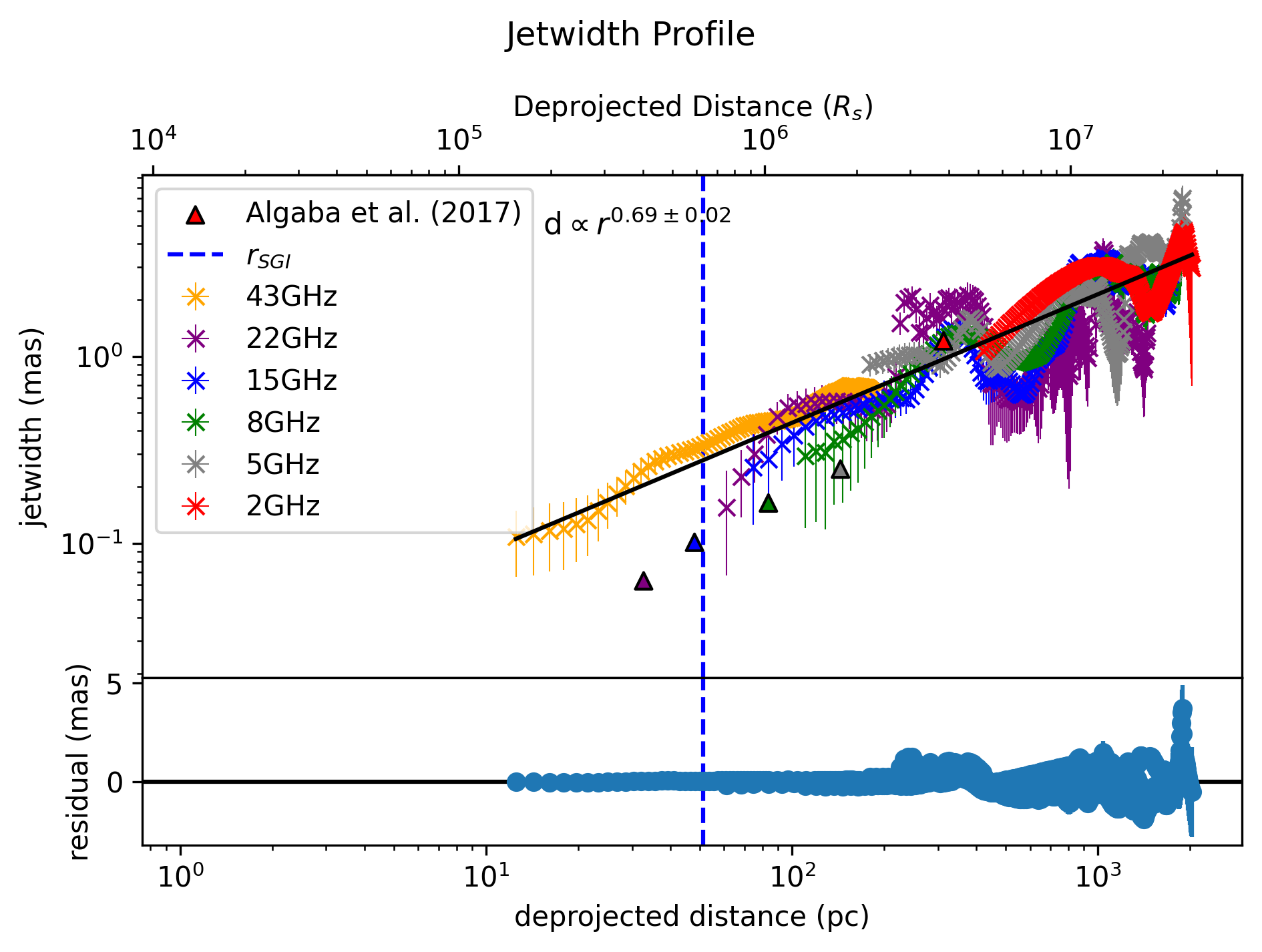}
    \caption{Top panel: Jet Width profile at multi-frequency. Bottom panel: residual plot. The vertical blue dotted line indicates the Sphere of Gravitational Influence (SGI). Dots indicate the core size from Algaba's work \citep{algaba2017resolving}. The colour correspond to the frequencies used in the plot}
    \label{fig:jetwidth}
\end{figure*}

\section{Results}
\subsection{Jet streamline}
To derive the jet streamline of CTA 102, which exhibits a southward jet orientation, we rotated the jet images for all frequencies by $60^\circ$ clockwise to align the jet visually parallel to the horizontal axis. Slices were taken perpendicular to the jet axis, starting from the core and spaced every two pixel out (correspond to one-fifth of the full width half maximum of the beam size) to obtain the flux density profile. A single Gaussian function was fitted to the profile to determine the local flux density maxima along each slice. These maxima were then connected to form a continuous streamline across the jet, which was subsequently smoothed using a third-order Savitzky-Golay filter \citep{savitzky1964smoothing}, mitigating local deviations and facilitating a robust comparison of mean jet widths \citep{paraschos2022jet}. 

To assess the influence of rotation angle on the derived streamline, the procedure was repeated across several rotation angles, ranging from $40{^\circ}-70{^\circ}$ with $5{^\circ}$ increment. Uncertainties arising from the Gaussian fitting, and rotation angles were quadratically added and calculated using the equation 3 in \citet{lee2008global}. A sample plot of the slicing method and resulting streamlines for 15 GHz data is shown in the Fig.~\ref{fig:streamline rotation}. As a consistency check, we also employed the method proposed by \citet{perucho2012anatomy}, which determines the geometrical center of the flux distribution along each slice. We found no significant differences with the previous Gaussian fitting method and thus we employed the initial Gaussian method for other frequencies. 

\subsection{Helical Jet Model}
\citet{fromm2013catchingIII} proposed that the jet of CTA 102 is best described by helical patterns or flow motion developing within an over-pressured jet in their paper. To investigate the helical pattern, we adopted the equation from \citet{algaba2019exploring} on the 15 GHz jet streamline. 
\begin{equation}
	{x^{\rm{proj}}_{\rm jet}}=F(u)\cos(\lambda)-G(u)\sin(\lambda)
\end{equation}
\begin{equation}
	{y^{\rm{proj}}_{\rm jet}}=F(u)\sin(\lambda)+G(u)\cos(\lambda)
\end{equation}
where
\begin{equation}
	{F(u)}=b u^{\epsilon}\cos(u-\phi)+ a u\sin(\theta)
\end{equation}
\begin{equation}
	{G(u)}=b u^{\epsilon}\sin(u-\phi)
\end{equation}
with u the polar angle measured in the plane perpendicular to the jet axis, $\phi$ an initial phase, $\lambda$ the P.A. rotation angle, the power law index $\epsilon=0.69$, and the viewing angle $\theta=2.6^\circ$. The plot is shown in Fig.~\ref{fig:helical_fit}. We may say the helical model can only fit the pattern from the streamline at around 10 mas, the amplitude of the oscillation is much larger than what is expected from the model. Therefore, although this cannot be ruled out, we consider that the fit does not reconstruct the jet streamline well. If the helical pattern is to be interpreted as shaping the streamline of the jet, as suggested by Fromm, this interpretation may not fully apply to CTA~102. 

\subsection{Jet Width Profile}
The jet width profile was derived by taking transverse slices perpendicular to the jet streamline obtained from previous section. These slices were spaced at intervals equivalent to one-fifth of the full width half maximum (FWHM) of the synthesized beam. Each flux density profile was fitted with a single Gaussian, and deconvolved with the defined circular beamsize, following the relation ${\theta_{\rm{jet}}}^{2}={\theta_{\rm{fit}}}^{2}-{\theta_{\rm{res}}}^{2}$ \citep{asada2012structure}. The defined beamsize for multi-epoch frequencies (15, 22, and 43 GHz) is the average beamsize taken from all epochs. The jet width was then parametrized as a power law $W_{\rm{jet}} \propto d^{\epsilon}$ where d is the deprojected distance. We excluded the ten innermost data points from the fitting due to their unresolved nature near the core, as well as the outermost data points where the SNR was insufficient.

The jet base or the core position is located at the optically thick region ($\tau = 1$), and its position is frequency dependent. In order to correctly combine the multifrequency map, we must apply the core-shift relative to a reference frequency using the equation $r_{\rm{core}}=A (\nu^{-1/k_{\rm{r}}}-\nu_{\rm{ref}}^{ -1/k_{\rm{r}}})$ \citep{blandford1979relativistic, lobanov1998ultracompact, o2009magnetic}. In our case, since the absolute position is lost in the calibration and imaging process, we adopted the lower and upper limit of A and $k_{\rm{r}}$ values calculated by \citet{fromm2013catchingIII} in their table 5, using 43 GHz as the reference frequency. We recognise that the core shift may vary across different observation dates. To account for this variability, we performed a Monte Carlo (MC) simulation, generating 500 core shifts realisations within the associated uncertainties and applying them to our case study. The resulting power-law index was taken computed as the average across all realisations, and the jet width profile corresponding to this average index is presented in Fig.~\ref{fig:jetwidth}.

\section{Discussion}
\subsection{Unseen transition in CTA 102}
The jet in CTA 102 analysed in this paper does not show a clear transition from parabolic to conical structure. Instead, a single power law fit to the jet width profile derived from the jet streamline, yields a quasi-parabolic shape with an index of $\epsilon = 0.69 \pm 0.02$, which is smaller than the results reported by \citet{fromm2013catchingIII} by 5 $\sigma$. This dicrepancy may be attributed to the differences in the rotation angle used to align the jet with the Relative Declination, inducing uncertainty. We also consider that a fraction of the discrepancy between our value and that from Fromm may arise from the core shift used in our MC simulation. For instance, if no core shift were applied, the power law index would be $\epsilon = 0.63 \pm 0.02$, which is different (around 2 $\sigma$) from the value obtained when considering the core shift.

To access the suitability of a single power law fit for our case, we examined the residuals of the fit alongside the jet width profile to determine if a clear pattern emerged around the Bondi radius, that could indicate a structural transtion. The residuals, however, do not show significant departures from zero, supporting the robustness of the fit and suggesting that no jet break is observed around the Bondi radius in CTA~102.

The obtained power law fit value, while distinct from both purely conical and parabolic geometries, lies within the quasi-parabolic regime. This observation is consistent with findings in other sources and has been statistically reported by \citet{algaba2017resolving} and theoretically explained by \citet{nakamura2018parabolic}.

For context, we compared our results with those from other well-studied jets, such as M87 \citep{asada2012structure}, which exhibits a parabolic shape upstream ($\epsilon \approx 0.58$) and a conical shape downstream ($\epsilon \approx 1.04$); NGC 6251 \citep{tseng2016structural}(parabolic upstream: $\epsilon \approx 0.5$ and conical downstream:$\epsilon \approx 1.06$), 3C 264 \citep{boccardi2019tev}($\epsilon = 0.40 \pm 0.04 $), and we find out that our value is much larger and more towards the conical regime. We also considered the recent work by \citet{kravchenko2025mojave} which suggests a break at radial distance 6.22 mas (approximately 137 mas deprojected, assuming a viewing angle of $2.6^\circ$) from the 15 GHz core. In our analysis, we observe fluctuations in the jet width around this region but do not find conclusive evidence supporting the presence of a distinct structural break.
 
In addition to our own data, we incorporate the core data from \citet{algaba2017resolving}. Interestingly, their works correlate with our 22 GHz data but deviate from the 43 GHz data, indicating a narrower jet width than predicted by our power-law fit. One possible explanation is that different frequencies probe different regions of the jet streamline as proposed by \citet{ghisellini1985inhomogeneous}. They suggests that the spectral shape may change from a partially opaque to transparent synchrotron emission, thereby changing streamline and jetwidth observed. A similar scenario has been suggested for M87 where \citet{asada2016indication} identified three streamlines coming from different region of the jet. However, we inspected the spectral index map for the source from \citep{fromm2013catchingIII} but did not find sufficient evidence to support the explanation. Another possible scenario could involve the long-term time variability, which may cause significant shifts in the location of the active inner region relative to the outer jet. This possibility is supported by Fig.~\ref{fig: singlevsstack} which shows discrepancies between the single and stack images in the inner jet region. 

Building on the concepts of Bondi radius and SGI introduced earlier, we estimated their respective scales for CTA~102 to examine potential correlations with features observed in the jet width profile. \citet{bednarek1994gamma} calculated the temperature $T_{\rm{max}}=3.5 \times10^6 K$ allows us to approximate the Bondi radius to $1.12 \times 10^6 R_{\rm{s}}$ (11.3 mas deprojected). Alternatively, by applying the M-$\sigma$ relation given by \citet{algaba2019exploring}, taking the equation $M_{\rm{BH}}=1.9\times 10^8 M_{\odot} (\sigma/[200\, \rm{km\, s^{-1}}])^{5.12}$, we calculated the stellar velocity dispersion value to be around 268.1 $\rm{km\,s}^{-1}$ and therefore estimated the SGI radius to $6.25\times 10^5 R_{\rm{s}}$ (corresponding to 6.32 mas). This value is roughly one order of magnitude smaller than the Bondi radius, but is consistent to the transition value by \citet{kovalev2020transition} ($r_{\rm{break}}$ within the range of $10^5$ to $10^6 R_{\rm{g}}$. 

Given the potential variability and uncertainty in the Bondi radius due to temporal changes in accretion flow and external conditions, we chose to focus our analysis on the SGI radius. This radius was overlaid on the jet width profile in Fig.~\ref{fig:jetwidth}. However, no distinct structural transition or significant peak in the jet width was observed near this location. We further explore the implications of this result and its relation to jet collimation mechanisms in the following section

\subsection{Possible Conditions Governing Jet Transition}
\begin{figure}
	\includegraphics[width=\columnwidth]{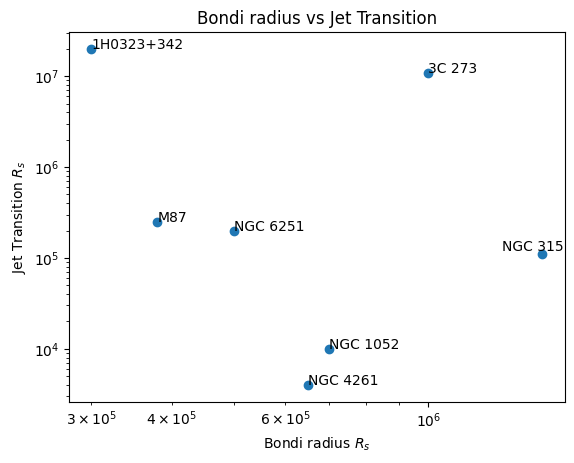}
    \caption{Jet Transition against Bondi Radius. 1H0303+342 \citep{kovalev2020transition}; M87 \citep{asada2012structure}; NGC 4261 \citep{nakahara2018finding}; NGC 315 \citep{ricci2022exploring}; 3C 273 \citep{okino2022collimation}; NGC 6251 \citep{tseng2016structural}; NGC 1052 \citep{baczko2022ambilateral}}
    \label{fig:Jet Transition}
\end{figure}

Based on our analysis of CTA~102, the absence of a jet break suggests that the location of the Bondi radius may not be the only constraint on the jet geometry in this source. Therefore, and in conjuction with previous studies on other sources, that also indicate either a lack of jet geometry break, or a jet geometry break hapenning at a totally different location, leads us to consider that other parameters may also play a crucial role. To make more quantitative assessment of these considerations, we present a plot of Bondi Radius vs Jet Transition Location for several sources in Fig.~\ref{fig:Jet Transition}. We calculated the coefficient of determination, $R^2$ value to be 0.0723, indicating that there is no clear correlation between these two parameters. While we acknowledge that there is a large uncertainty in these values, possibly of the order of an order of magnitude, we can still qualitatively discuss the Bondi radius with the jet transition. We infer that Bondi radius may not be the main parameter give rise to the transition, and may be affected by other phenomenology such as the magnetic field structure around the central black hole.

The common discussion on the topic of the magnetic field can be separated into two models, the SANE (Standard and Normal Evolution) and MAD (Magnetically arrested disk) \citep{igumenshchev2003three, igumenshchev2008magnetically, narayan2003magnetically, mckinney2012general}. The SANE model ejecting the angular momentum outwards via turbulence driven by the magneto-rotational instability whereas the MAD model creates discrete 'blobs' of material outwards due to disruption of the axisymmetric accretion flow by the strong magnetic field. \citet{yuan2022accretion}, who investigated the rotation measure (RM) of M87 found that the values predicted by the MAD model are consistent with the observations, whereas the SANE model overestimates it by two orders of magnitude. \citet{ricci2022exploring} also found a good agreement between a MAD model with the quasi-parabolic jet in NGC 315. In spite of its importance to understand the effect of magnetic field on the jet transition, it has been difficult to observe the magnetic field at the jet launching region due to very weak polarization at these regions. With the recent flourish development in the higher frequency observation using EHT, it would be a great help for us to probe further upstream into the black hole-jet connecting region. 

\subsection{Jet Humps}
Several humps were identified in the jet width profile of CTA~102, situated around 350 and 1000 pc deprojected, correspond to $4.3\times 10^6 R_{\rm{s}}$ and $ 1.2\times 10^7 R_{\rm{s}}$ respectively, as seen in Fig.~\ref{fig:jetwidth}. Notably, these features are distinct from the observed jet bending region, located farther downstream at nearly 13 mas projected (approximately 2324 pc deprojected or $2.9\times 10^7 R_{\rm{s}}$), as seen in Fig.~\ref{fig: All Clean Map}. This pattern has also been noted in the MOJAVE survey jet width profiles \citep{pushkarev2017mojave}. Likewise, \citet{kravchenko2025mojave} described a "zig-zag" morphology in certain sources, notably 1716+686, which may be analogous to the jet hump phenomenon discussed here.

To qualitatively investigate potential correlations between these features across AGN classification, we select the blazars BL Lac \citep{casadio2021jet} and 3C~273 \citep{okino2022collimation}, and the radio galaxies M87 \citep{nikonov2023properties} and 3C~264 \citep{boccardi2019tev}. We extract the jet width profile data using WebPlotDigitizer (Version 4 \footnote {\url{https://apps.automeris.io/wpd4/}} \citet{marin2017webplotdigitizer}), a tool validated in previous literature \citep{drevon2017intercoder} as a reliable method for digitizing data from figures. The data for 3C~264, M87 and 3C~273 are retrieved from their respective publications, and the data for BL Lac directly from C. Casadio (personal communication).

\begin{figure}
	\includegraphics[width=\columnwidth]{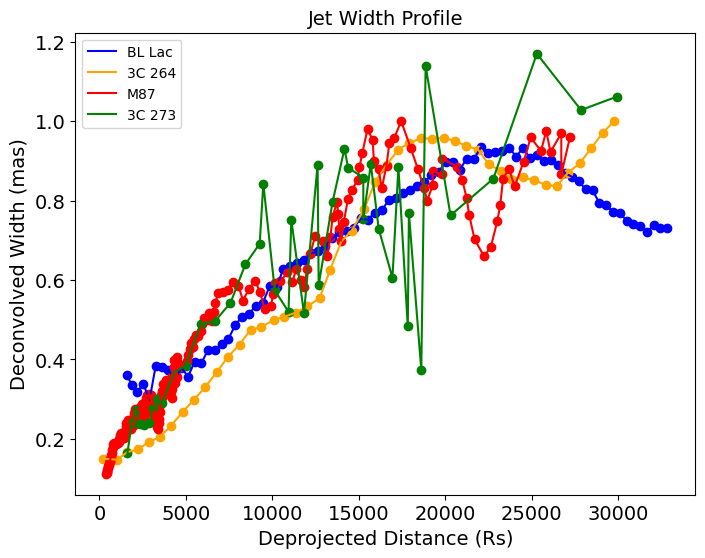}
    \caption{Jet Width Profile of M87, 3C~273, 3C~264 and BL Lac}
    \label{fig:jetwidth_other}
\end{figure}

To enable direct comparison across these sources that have different intrinsic opening angles, we normalise the jet width profiles. The resulting comparison, shown in Fig.~\ref{fig:jetwidth_other}, reveals a clustering of humps around deprojected distance of $1.5 \times 10^4 R_{\rm{s}}$ to $2.5\times 10^4 R_{\rm{s}}$, suggesting a structural resemblance among these sources. 

While the physical origin of these humps is not to be fully discussed here, we tentatively propose the occurrence of this characteristic to the quad relativistic magnetohydrodynamic shock fronts mentioned in \citet{nakamura2010magnetohydrodynamic}. The forward and reverse shocks may be an indication of peak observing in the jetwidth profile in relation with the current-driven (CD) kink instability that disrupts the magnetic field structure, leading multiple standing recollimation shocks in the jet. 

This apparent consistency, irrespective of AGN type, leads us to propose that such jet width humps may represent a recurring feature in relativistic jet. A comprehensive statistical analysis of these features across a broader sample of AGN is beyond the scope of the present work. Further analysis will be done to confirm the proposal. 

\section{Conclusions}
In this work, we analyzed the jet width profile of CTA~102 and found a quasi-parabolic jet structure ($\epsilon = 0.69 \pm 0.02 $). Our result indicated no clear transition found in the jet width profile of CTA~102, which contrasts with the recent findings found by \citet{kravchenko2025mojave}, who proposed a jet recollimation at around 7 mas radial distance (154 mas deprojected) from the 15 GHz core. Furthermore, we could not reconstruct a helical jet structure as suggested by \citet{fromm2013catchingII}. We extended our analysis by comparing the location of jet transitions and Bondi radii across several sources and found no strong correlation between these parameters. This suggests that the Bondi radius alone is insufficient to explain the jet collimation behaviour. Additional factors, such as the magnetic field structure of the jet particularly the MAD model proved to have good agreement relating jet transition in several sources. Additionally, we observed the existence of several "humps" in the jet width profile of CTA~102, and performed a qualitative comparison with the similar features seen in other sources, namely 3C~264, M87, 3C~273, and BL Lac. Our findings lead us to propose that the jet hump may be a recurring structural feature in a subset of AGN jets. Further investigation utilising the MOJAVE data is planned to explore this feature in greater detail.

\section*{Acknowledgements}
This research made use of the archive data from the National Radio Astronomy Observatory (NRAO), VLBA-BU Blazar Monitoring Program (BEAM-ME and VLBA-BU-BLAZAR; http://www.bu.edu/blazars/BEAM-ME.html), funded by NASA through the Fermi Guest Investigator Program. The VLBA is an instrument of the National Radio Astronomy Observatory. The National Radio Astronomy Observatory is a facility of the National Science Foundation operated by Associated Universities, Inc. The work also includes stacked image from MOJAVE data base, mainted by the MOJAVE team \citep{lister2009mojave}. We gratefully acknowledge Carolina Casadio, Kazunori Akiyama and Hiroki Okino for generously granting permission to use the jet width data for BL Lac and 3C 273 respectively. We also thank Kah Chun Lai, Shoko Koyama, Keichi Asada and the anonymous referee for the useful discussion, which highly improved the quality of the paper. 

\section{Data Availability}
The data supporting this article can be accessed from the NRAO archive (\url{https://science.nrao.edu/}) under project codes BA064, V019, and BW086; the 15 GHz stacked data are available from the MOJAVE data archive (\url{https://www.cv.nrao.edu/MOJAVE/}), and the 43 GHz stacked data can be found at BU program (\url{https://www.bu.edu/blazars/})

\bibliographystyle{mnras}
\bibliography{references} 

\begin{thebibliography}{}
\makeatletter
\relax
\def\mn@urlcharsother{\let\do\@makeother \do\$\do\&\do\#\do\^\do\_\do\%\do\~}
\def\mn@doi{\begingroup\mn@urlcharsother \@ifnextchar [ {\mn@doi@}
  {\mn@doi@[]}}
\def\mn@doi@[#1]#2{\def\@tempa{#1}\ifx\@tempa\@empty \href
  {http://dx.doi.org/#2} {doi:#2}\else \href {http://dx.doi.org/#2} {#1}\fi
  \endgroup}
\def\mn@eprint#1#2{\mn@eprint@#1:#2::\@nil}
\def\mn@eprint@arXiv#1{\href {http://arxiv.org/abs/#1} {{\tt arXiv:#1}}}
\def\mn@eprint@dblp#1{\href {http://dblp.uni-trier.de/rec/bibtex/#1.xml}
  {dblp:#1}}
\def\mn@eprint@#1:#2:#3:#4\@nil{\def\@tempa {#1}\def\@tempb {#2}\def\@tempc
  {#3}\ifx \@tempc \@empty \let \@tempc \@tempb \let \@tempb \@tempa \fi \ifx
  \@tempb \@empty \def\@tempb {arXiv}\fi \@ifundefined
  {mn@eprint@\@tempb}{\@tempb:\@tempc}{\expandafter \expandafter \csname
  mn@eprint@\@tempb\endcsname \expandafter{\@tempc}}}

\bibitem[\protect\citeauthoryear{Algaba, Nakamura, Asada  \& Lee}{Algaba
  et~al.}{2017}]{algaba2017resolving}
Algaba J.-C.,  Nakamura M.,  Asada K.,   Lee S.-S.,  2017, The Astrophysical
  Journal, 834, 65

\bibitem[\protect\citeauthoryear{Algaba, Rani, Lee, Kino, Park  \& Kim}{Algaba
  et~al.}{2019}]{algaba2019exploring}
Algaba J.~C.,  Rani B.,  Lee S.-S.,  Kino M.,  Park J.,   Kim J.-Y.,  2019, The
  Astrophysical Journal, 886, 85

\bibitem[\protect\citeauthoryear{Asada \& Nakamura}{Asada \&
  Nakamura}{2012}]{asada2012structure}
Asada K.,  Nakamura M.,  2012, The Astrophysical Journal Letters, 745, L28

\bibitem[\protect\citeauthoryear{Asada, Nakamura  \& Pu}{Asada
  et~al.}{2016}]{asada2016indication}
Asada K.,  Nakamura M.,   Pu H.-Y.,  2016, The Astrophysical Journal, 833, 56

\bibitem[\protect\citeauthoryear{Baczko et~al.,}{Baczko
  et~al.}{2022}]{baczko2022ambilateral}
Baczko A.-K.,  et~al., 2022, Astronomy \& Astrophysics, 658, A119

\bibitem[\protect\citeauthoryear{Bednarek \& Kirk}{Bednarek \&
  Kirk}{1994}]{bednarek1994gamma}
Bednarek W.,  Kirk J.,  1994, arXiv preprint astro-ph/9408013

\bibitem[\protect\citeauthoryear{Blandford \& K{\"o}nigl}{Blandford \&
  K{\"o}nigl}{1979}]{blandford1979relativistic}
Blandford R.,  K{\"o}nigl A.,  1979, Astrophysical Journal, Part 1, vol. 232,
  Aug. 15, 1979, p. 34-48., 232, 34

\bibitem[\protect\citeauthoryear{Blandford \& Payne}{Blandford \&
  Payne}{1982}]{blandford1982hydromagnetic}
Blandford R.~D.,  Payne D.,  1982, Monthly Notices of the Royal Astronomical
  Society, 199, 883

\bibitem[\protect\citeauthoryear{Blandford \& Znajek}{Blandford \&
  Znajek}{1977}]{blandford1977electromagnetic}
Blandford R.~D.,  Znajek R.~L.,  1977, Monthly Notices of the Royal
  Astronomical Society, 179, 433

\bibitem[\protect\citeauthoryear{Boccardi, Migliori, Grandi, Torresi, Mertens,
  Karamanavis, Angioni  \& Vignali}{Boccardi et~al.}{2019}]{boccardi2019tev}
Boccardi B.,  Migliori G.,  Grandi P.,  Torresi E.,  Mertens F.,  Karamanavis
  V.,  Angioni R.,   Vignali C.,  2019, Astronomy \& Astrophysics, 627, A89

\bibitem[\protect\citeauthoryear{Casadio et~al.,}{Casadio
  et~al.}{2021}]{casadio2021jet}
Casadio C.,  et~al., 2021, Astronomy \& Astrophysics, 649, A153

\bibitem[\protect\citeauthoryear{Drevon, Fursa  \& Malcolm}{Drevon
  et~al.}{2017}]{drevon2017intercoder}
Drevon D.,  Fursa S.~R.,   Malcolm A.~L.,  2017, Behavior modification, 41, 323

\bibitem[\protect\citeauthoryear{Fromm et~al.,}{Fromm
  et~al.}{2011}]{fromm2011catchingI}
Fromm C.~M.,  et~al., 2011, Astronomy \& Astrophysics, 531, A95

\bibitem[\protect\citeauthoryear{Fromm et~al.,}{Fromm
  et~al.}{2013a}]{fromm2013catchingII}
Fromm C.~M.,  et~al., 2013a, Astronomy \& Astrophysics, 551, A32

\bibitem[\protect\citeauthoryear{Fromm, Ros, Perucho, Savolainen, Mimica,
  Kadler, Lobanov  \& Zensus}{Fromm et~al.}{2013b}]{fromm2013catchingIII}
Fromm C.,  Ros E.,  Perucho M.,  Savolainen T.,  Mimica P.,  Kadler M.,
  Lobanov A.,   Zensus J.,  2013b, Astronomy \& Astrophysics, 557, A105

\bibitem[\protect\citeauthoryear{Fromm, Perucho, Ros, Savolainen  \&
  Zensus}{Fromm et~al.}{2015}]{fromm2015location}
Fromm C.~M.,  Perucho M.,  Ros E.,  Savolainen T.,   Zensus J.~A.,  2015,
  Astronomy \& Astrophysics, 576, A43

\bibitem[\protect\citeauthoryear{Ghisellini, Maraschi  \& Treves}{Ghisellini
  et~al.}{1985}]{ghisellini1985inhomogeneous}
Ghisellini G.,  Maraschi L.,   Treves A.,  1985, Astronomy and Astrophysics
  (ISSN 0004-6361), vol. 146, no. 2, May 1985, p. 204-212., 146, 204

\bibitem[\protect\citeauthoryear{Greenstein \& Schmidt}{Greenstein \&
  Schmidt}{1979}]{greenstein1979quasi}
Greenstein J.~L.,  Schmidt M.,  1979, in , A Source Book in Astronomy and
  Astrophysics, 1900--1975.
Harvard University Press, pp 811--818

\bibitem[\protect\citeauthoryear{Harris \& Roberts}{Harris \&
  Roberts}{1960}]{harris1960radio}
Harris D.,  Roberts J.,  1960, Publications of the Astronomical Society of the
  Pacific, 72, 237

\bibitem[\protect\citeauthoryear{Igumenshchev}{Igumenshchev}{2008}]{igumenshchev2008magnetically}
Igumenshchev I.~V.,  2008, The Astrophysical Journal, 677, 317

\bibitem[\protect\citeauthoryear{Igumenshchev, Narayan  \&
  Abramowicz}{Igumenshchev et~al.}{2003}]{igumenshchev2003three}
Igumenshchev I.~V.,  Narayan R.,   Abramowicz M.~A.,  2003, The Astrophysical
  Journal, 592, 1042

\bibitem[\protect\citeauthoryear{Jorstad et~al.,}{Jorstad
  et~al.}{2005}]{jorstad2005polarimetric}
Jorstad S.~G.,  et~al., 2005, The Astronomical Journal, 130, 1418

\bibitem[\protect\citeauthoryear{Kellermann, Sramek, Schmidt, Shaffer  \&
  Green}{Kellermann et~al.}{1989}]{kellermann1989vla}
Kellermann K.,  Sramek R.,  Schmidt M.,  Shaffer D.,   Green R.,  1989,
  Astronomical Journal (ISSN 0004-6256), vol. 98, Oct. 1989, p. 1195-1207., 98,
  1195

\bibitem[\protect\citeauthoryear{Kovalev, Pushkarev, Nokhrina, Plavin, Beskin,
  Chernoglazov, Lister  \& Savolainen}{Kovalev
  et~al.}{2020}]{kovalev2020transition}
Kovalev Y.~Y.,  Pushkarev A.~B.,  Nokhrina E.~E.,  Plavin A.~V.,  Beskin V.~S.,
   Chernoglazov A.~V.,  Lister M.~L.,   Savolainen T.,  2020, Monthly Notices
  of the Royal Astronomical Society, 495, 3576

\bibitem[\protect\citeauthoryear{Kravchenko, Pashchenko, Homan, Kovalev,
  Lister, Pushkarev, Ros  \& Savolainen}{Kravchenko
  et~al.}{2025}]{kravchenko2025mojave}
Kravchenko E.,  Pashchenko I.,  Homan D.,  Kovalev Y.,  Lister M.,  Pushkarev
  A.,  Ros E.,   Savolainen T.,  2025, Monthly Notices of the Royal
  Astronomical Society, 538, 2008

\bibitem[\protect\citeauthoryear{Lee, Lobanov, Krichbaum, Witzel, Zensus,
  Bremer, Greve  \& Grewing}{Lee et~al.}{2008}]{lee2008global}
Lee S.-S.,  Lobanov A.~P.,  Krichbaum T.~P.,  Witzel A.,  Zensus A.,  Bremer
  M.,  Greve A.,   Grewing M.,  2008, The Astronomical Journal, 136, 159

\bibitem[\protect\citeauthoryear{Lister et~al.,}{Lister
  et~al.}{2009}]{lister2009mojave}
Lister M.,  et~al., 2009, The Astronomical Journal, 137, 3718

\bibitem[\protect\citeauthoryear{Lobanov}{Lobanov}{1998}]{lobanov1998ultracompact}
Lobanov A.,  1998, Astronomy and Astrophysics, v. 330, p. 79-89 (1998), 330, 79

\bibitem[\protect\citeauthoryear{Marin, Rohatgi  \& Charlot}{Marin
  et~al.}{2017}]{marin2017webplotdigitizer}
Marin F.,  Rohatgi A.,   Charlot S.,  2017, arXiv preprint arXiv:1708.02025

\bibitem[\protect\citeauthoryear{McKinney, Tchekhovskoy  \& Blandford}{McKinney
  et~al.}{2012}]{mckinney2012general}
McKinney J.~C.,  Tchekhovskoy A.,   Blandford R.~D.,  2012, Monthly Notices of
  the Royal Astronomical Society, 423, 3083

\bibitem[\protect\citeauthoryear{Nakahara, Doi, Murata, Hada, Nakamura  \&
  Asada}{Nakahara et~al.}{2018}]{nakahara2018finding}
Nakahara S.,  Doi A.,  Murata Y.,  Hada K.,  Nakamura M.,   Asada K.,  2018,
  The Astrophysical Journal, 854, 148

\bibitem[\protect\citeauthoryear{Nakahara, Doi, Murata, Nakamura, Hada, Asada,
  Sawada-Satoh  \& Kameno}{Nakahara et~al.}{2019}]{nakahara2019two}
Nakahara S.,  Doi A.,  Murata Y.,  Nakamura M.,  Hada K.,  Asada K.,
  Sawada-Satoh S.,   Kameno S.,  2019, The Astronomical Journal, 159, 14

\bibitem[\protect\citeauthoryear{Nakamura \& Asada}{Nakamura \&
  Asada}{2013}]{nakamura2013parabolic}
Nakamura M.,  Asada K.,  2013, The Astrophysical Journal, 775, 118

\bibitem[\protect\citeauthoryear{Nakamura, Garofalo  \& Meier}{Nakamura
  et~al.}{2010}]{nakamura2010magnetohydrodynamic}
Nakamura M.,  Garofalo D.,   Meier D.~L.,  2010, The Astrophysical Journal,
  721, 1783

\bibitem[\protect\citeauthoryear{Nakamura et~al.,}{Nakamura
  et~al.}{2018}]{nakamura2018parabolic}
Nakamura M.,  et~al., 2018, The Astrophysical Journal, 868, 146

\bibitem[\protect\citeauthoryear{Narayan, Igumenshchev  \& Abramowicz}{Narayan
  et~al.}{2003}]{narayan2003magnetically}
Narayan R.,  Igumenshchev I.~V.,   Abramowicz M.~A.,  2003, Publications of the
  Astronomical Society of Japan, 55, L69

\bibitem[\protect\citeauthoryear{Nikonov, Kovalev, Kravchenko, Pashchenko  \&
  Lobanov}{Nikonov et~al.}{2023}]{nikonov2023properties}
Nikonov A.,  Kovalev Y.,  Kravchenko E.,  Pashchenko I.,   Lobanov A.,  2023,
  Monthly Notices of the Royal Astronomical Society, 526, 5949

\bibitem[\protect\citeauthoryear{O'Sullivan \& Gabuzda}{O'Sullivan \&
  Gabuzda}{2009}]{o2009magnetic}
O'Sullivan S.~P.,  Gabuzda D.~C.,  2009, Monthly Notices of the Royal
  Astronomical Society, 400, 26

\bibitem[\protect\citeauthoryear{Okino et~al.,}{Okino
  et~al.}{2022}]{okino2022collimation}
Okino H.,  et~al., 2022, The Astrophysical Journal, 940, 65

\bibitem[\protect\citeauthoryear{Paraschos et~al.,}{Paraschos
  et~al.}{2022}]{paraschos2022jet}
Paraschos G.,  et~al., 2022, Astronomy \& Astrophysics, 665, A1

\bibitem[\protect\citeauthoryear{Perucho, Kovalev, Lobanov, Hardee  \&
  Agudo}{Perucho et~al.}{2012}]{perucho2012anatomy}
Perucho M.,  Kovalev Y.,  Lobanov A.,  Hardee P.,   Agudo I.,  2012, The
  Astrophysical Journal, 749, 55

\bibitem[\protect\citeauthoryear{Pushkarev, Kovalev, Lister  \&
  Savolainen}{Pushkarev et~al.}{2017}]{pushkarev2017mojave}
Pushkarev A.~B.,  Kovalev Y.,  Lister M.,   Savolainen T.,  2017, Monthly
  Notices of the Royal Astronomical Society, 468, 4992

\bibitem[\protect\citeauthoryear{Ricci et~al.,}{Ricci
  et~al.}{2022}]{ricci2022exploring}
Ricci L.,  et~al., 2022, Astronomy \& Astrophysics, 664, A166

\bibitem[\protect\citeauthoryear{Russell, Fabian, McNamara  \&
  Broderick}{Russell et~al.}{2015}]{russell2015inside}
Russell H.,  Fabian A.,  McNamara B.,   Broderick A.,  2015, Monthly Notices of
  the Royal Astronomical Society, 451, 588

\bibitem[\protect\citeauthoryear{Savitzky \& Golay}{Savitzky \&
  Golay}{1964}]{savitzky1964smoothing}
Savitzky A.,  Golay M.~J.,  1964, Analytical chemistry, 36, 1627

\bibitem[\protect\citeauthoryear{Schmidt}{Schmidt}{1965}]{schmidt1965large}
Schmidt M.,  1965, Astrophysical Journal, vol. 141, p. 1295, 141, 1295

\bibitem[\protect\citeauthoryear{Shepherd}{Shepherd}{1997}]{shepherd1997difmap}
Shepherd M.,  1997, in Astronomical Data Analysis Software and Systems VI.
  p.~77

\bibitem[\protect\citeauthoryear{Strittmatter, Hill, Pauliny-Toth, Steppe  \&
  Witzel}{Strittmatter et~al.}{1980}]{strittmatter1980radio}
Strittmatter P.,  Hill P.,  Pauliny-Toth I.,  Steppe H.,   Witzel A.,  1980,
  Astronomy and Astrophysics, vol. 88, no. 3, Aug. 1980, p. L12-L15., 88, L12

\bibitem[\protect\citeauthoryear{Tseng, Asada, Nakamura, Pu, Algaba  \&
  Lo}{Tseng et~al.}{2016}]{tseng2016structural}
Tseng C.-Y.,  Asada K.,  Nakamura M.,  Pu H.-Y.,  Algaba J.-C.,   Lo W.-P.,
  2016, The Astrophysical Journal, 833, 288

\bibitem[\protect\citeauthoryear{Urry \& Padovani}{Urry \&
  Padovani}{1995}]{urry1995unified}
Urry C.~M.,  Padovani P.,  1995, Publications of the Astronomical Society of
  the Pacific, 107, 803

\bibitem[\protect\citeauthoryear{V{\'e}ron-Cetty \& V{\'e}ron}{V{\'e}ron-Cetty
  \& V{\'e}ron}{2003}]{veron2003catalogue}
V{\'e}ron-Cetty M.-P.,  V{\'e}ron P.,  2003, Astronomy \& Astrophysics, 412,
  399

\bibitem[\protect\citeauthoryear{Wells}{Wells}{1985}]{wells1985nrao}
Wells D.,  1985, in , Data Analysis in Astronomy.
Springer, pp 195--209

\bibitem[\protect\citeauthoryear{Xiao, Zhu, Fu, Zhang  \& Fan}{Xiao
  et~al.}{2022}]{xiao2022radio}
Xiao H.,  Zhu J.,  Fu L.,  Zhang S.,   Fan J.,  2022, Publications of the
  Astronomical Society of Japan, 74, 239

\bibitem[\protect\citeauthoryear{Yan, Lu, Jiang, Krichbaum  \& Shen}{Yan
  et~al.}{2023}]{yan2023kinematics}
Yan X.,  Lu R.-S.,  Jiang W.,  Krichbaum T.~P.,   Shen Z.-Q.,  2023, The
  Astrophysical Journal, 957, 32

\bibitem[\protect\citeauthoryear{Yuan, Wang  \& Yang}{Yuan
  et~al.}{2022}]{yuan2022accretion}
Yuan F.,  Wang H.,   Yang H.,  2022, The Astrophysical Journal, 924, 124

\bibitem[\protect\citeauthoryear{Zamaninasab, Savolainen, Clausen-Brown,
  Hovatta, Lister, Krichbaum, Kovalev  \& Pushkarev}{Zamaninasab
  et~al.}{2013}]{zamaninasab2013evidence}
Zamaninasab M.,  Savolainen T.,  Clausen-Brown E.,  Hovatta T.,  Lister M.,
  Krichbaum T.,  Kovalev Y.,   Pushkarev A.,  2013, Monthly Notices of the
  Royal Astronomical Society, 436, 3341

\makeatother
\end{thebibliography}
\bsp	


\label{lastpage}
\end{document}